\documentstyle[epsf]{article}
\date{}
\begin{document}
{\bf UNCERTAINTY RELATIONS FOR ENTANGLED STATES}
\\
\vspace{.3in}
\begin{quote}
{\bf G. Rigolin}
\end{quote}
\begin{quote}
{\em Departamento de Raios C\'osmicos e Cronologia \\
Instituto de F\'{\i}sica Gleb Wataghin \\
Universidade Estadual de Campinas, C.P. 6165, cep 13084-971 \\
Campinas, S\~ao Paulo, Brazil\\
E-mail: rigolin@ifi.unicamp.br}
\end{quote}
\vspace{.6in}
A generalized uncertainty relation for an entangled pair of particles 
is obtained if we impose a symmetrization rule for all operators that we should 
employ when doing any calculation using the entangled wave function of the pair. 
This new relation reduces to Heisenberg's uncertainty relation when the particles 
have no correlation and suggests that we can have new lower bounds for the product 
of position and momentum dispersions.

\section{INTRODUCTION}

In this letter we examine the derivation of the uncertainty relations for a pair of entangled
particles motivated by the recent experiment of Kim and Shih [1], a realization of 
Popper's experiment [2], which can be considered an extension of the EPR argument [3].

Let us begin with a brief review of Popper's experiment, as done by Kim and Shih:
The entangled pair of photons are produced by Spontaneous Parametric Down Conversion (SPDC). 
Kim and Shih's measurements are conditional in the sense that the detection of photon 2 by a 
y-scanning detector $D_{2}$ is coincidental with the detection, in detector $D_{1}$, of photon 1
after its passage through slit A. (See figure 1). Two cases are considered:

\begin{enumerate}

\item[(a)] Slit A and slit B (placed along the trajectory of photon 2) have the same width.
\item[(b)] Slit B is left wide open.

\end{enumerate}

\begin{center}
\centering
\begin{figure}[ht]
\centering
\epsfxsize=4.0in
\centering
\epsffile{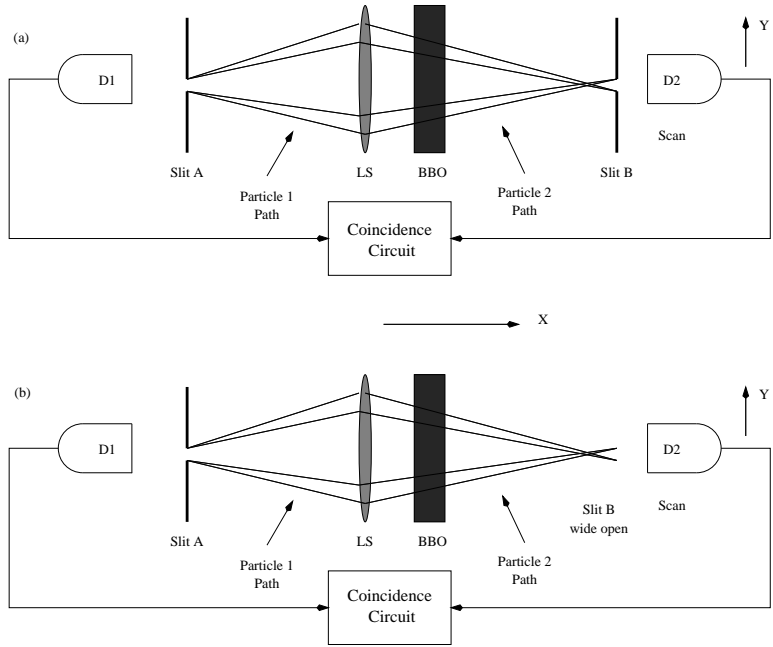}
\caption{ \footnotesize This is almost the same figure given in Kim and Shih's paper. Part (a) represents
the set-up with both slits with the same width. Part (b) represents slit B wide open. 
Beta Barium Borate (BBO) is the crystal where a laser beam produces by SPDC the 
entangled pair of photons. LS is the lens that produces a ``ghost image'' of slit A 
and localizes photon 2 when detecting photon 1 at slit A. See reference [1] for more
details.} 
\end{figure}
\end{center}
The first case does not present any challenge to our understanding as the Heisenberg 
uncertainty relation applies to both branches. The interesting situation comes from 
the second set-up, where Kim and Shih's experiment suggests that $\Delta y_{2}\Delta p_{y_{2}}<\hbar$ 
in an apparent violation of Heisenberg relation. They explain the result invoking the 
necessity of working with the entangled wave function of the pair (biphoton wavefunction)[1].

Two recent papers [4, 5] have discussed this problem from yet two different
 points of view. The paper by Short [4] claims that there is no violation of the uncertainty 
 principle and justifies this claim by affirming that the two photons:
 
\begin{quote}

\em do not interact with each other after their initial creation and must evolve  
independently between measurements when they are space-like separated.

\end{quote} 

We do not agree with this assumption because we think an entangled pair of photons need not obey this. 
Short's main argument is that the observed coincident patterns are dominated by a blurring of the photons' 
path which he considers intrinsic to the experimental set-up of Kim and Shih. However, we could expand the 
pump beam diameter and invalidate Short's analysis.

The second paper, by Unnikrishnan [5], approaches the problem in what seems to us 
the right but still incomplete way. Looking at the wave function of the entangled pair, 
 without showing any of the actual calculations, Unnikrishnan claims that the constraint of
 momentum conservation explains Kim and Shih results. In the following we give a complete 
 treatment of the problem calling attention to points that have passed unnoticed in the above 
  mentioned papers.
 
\section{OUR RESULTS}

In agreement with [1,5] we recognize that when dealing with entangled systems
 we should not use wave functions that describe the isolated evolution of a member of the 
 system, but rather we should use the entangled wave function of the system. These entangled
  wave functions have to obey the symmetrization rules of Quantum Mechanics:
(anti-)~symmetric wave functions for (fermions)~bosons.

It is however less known, despite appearing in some text books [6], the fact that for
 a correlated system (entangled system) we must deal with what is called
  {\em physical observables}, which have to obey symmetry requirements as well.
   As shown by Cohen-Tannoudji [6], {\em physical observables} must commute with all the
    permutation operators that appear in the system.

We restrict ourselves to the case of a pair of correlated particles, 
but a generalization to a system of N entangled particles is straightforward.

Let us define the following operator:
\begin{equation}
	{\cal O}(1,2)= \sum_{i=1}^{n}{A_{i}(1) \otimes B_{i}(2)},
\end{equation}
where $n$ is an integer greater than zero, not necessarily equal to the number of entangled particles 
(an example is the total angular momentum of two particles $J(1,2)=L(1) \otimes {\cal I}_{2} + S(1) 
\otimes {\cal I}_{2} + {\cal I}_{1} \otimes L(2) + {\cal I}_{1} \otimes S(2)$, where $L(i)$, $S(i)$ and 
${\cal I}_{i}$ are the orbital, spin angular momentum and the identity operator of particle $i$).
 $A_{i}(1)$ and $B_{i}(2)$ can be any observables 
initially defined in the state spaces ${\cal E}(1)$ and ${\cal E}(2)$ of particles 1 and 2, 
and then extended into ${\cal E}(1,2)$, the state space of the two-particle system. The state
space ${\cal E}(1,2)$ is the tensor product of the state spaces of particles 1 and 2,  
${\cal E}(1,2) = {\cal E}(1) \otimes {\cal E}(2)$.

The operator ${\cal O}(1,2)$ is called a {\em physical observable} if it satisfies the following 
commutation relation:
\begin{equation}
	[ {\cal O}(1,2), P_{21} ]=0,
\end{equation}
where $P_{21}$ is the permutation operator in the state space ${\cal E}(1,2)$. It can be shown [6] 
that $P_{21}$ is hermitian and obeys the following relation:
\begin{equation}
	P_{21}{\cal O}(1,2) P_{21}^{\dag}={\cal O}(2,1).  \label{adaga}
\end{equation}
Using Eq.~(\ref{adaga}) and defining the extended position and momentum operators, in a given direction,
 $Q(1,2)=Q(1) \otimes {\cal I}_{2} + {\cal I}_{1} \otimes Q(2)$ and
  $P(1,2)=P(1) \otimes {\cal I}_{2} + {\cal I}_{1} \otimes P(2)$, where ${\cal I}_{i}$ is the
  identity operator in the state space of particle {\em i}, we can show that:
\begin{equation}
	[ Q(1,2), P_{21} ]=[ P(1,2), P_{21} ]=0.
\end{equation}
Therefore, $Q(1,2)$ and $P(1,2)$ are {\em physical observables}.
 
Experimentally, in a coincidence measurement, $Q(1,2)$ is the sum 
of the positions of both particles and $P(1,2)$ gives the total 
momentum of the system in a given direction.
We now require that when deducing the uncertainty relation for a correlated system we should 
use only {\em physical observables} as follows:
\begin{equation}
	(\Delta Q(1,2))^{2}(\Delta P(1,2))^{2} \geq 
	\frac{{\mid \left< [Q(1,2),P(1,2)] \right> \mid}^{2}}{4}.  
	\label{geral}
\end{equation}
We cannot have the traditional relations
\begin{equation}
      (\Delta Q(i))^{2}(\Delta P(i))^{2} \geq 
      \frac{{\mid \left< [Q(i),P(i)] \right> \mid}^{2}}{4},
\end{equation}
because $Q(i)$ and $P(i)$, where $i=1$ or $i=2$, are not {\em physical observables} 
(they do not commute with the permutation operator). Manipulating Eq.~(\ref{geral}) we have 
(from now on we write $Q(i)=Q_{i}$ and $P(i)=P_{i}$ to simplify notation):
\begin{displaymath}
	[(\Delta Q_{1})^{2}+(\Delta Q_{2})^{2}+2(\left< Q_{1}Q_{2} \right> -
	 \left< Q_{1} \right> \left< Q_{2} \right>)] \times 
\end{displaymath}
\begin{equation}
	 [(\Delta P_{1})^{2}+(\Delta P_{2})^{2}+2(\left< P_{1}P_{2} \right>- 
	 \left< P_{1} \right> \left< P_{2} \right>)]  
	 \geq {\hbar}^{2}. 
\end{equation}
Assuming, as done by Popper and implictly by Kim and Shih, that $\Delta Q_{1} = \Delta Q_{2}$ we get:
\begin{displaymath}
	\left[ (\Delta Q_{2})^{2}+(\left< Q_{1}Q_{2} \right> - 
	\left< Q_{1} \right> \left< Q_{2} \right>) \right] \times 
\end{displaymath}
\begin{equation}
	\left[ \frac{(\Delta P_{1})^{2}}{2}+\frac{(\Delta P_{2})^{2}}{2}+
	(\left< P_{1}P_{2} \right> - \left< P_{1} \right> \left< P_{2} \right>) \right] 
	\geq \frac{{\hbar}^{2}}{4}. \label{eq39}
\end{equation}

This last expression should be the correct uncertainty relation when treating a correlated pair 
of particles and not the naive Heisenberg uncertainty relation:
\begin{equation}
	(\Delta Q_{2})^{2}(\Delta P_{2})^{2} \geq \frac{{\hbar}^{2}}{4}.
	 \label{velha}
\end{equation}
We suggest that Kim and Shih's experimental results should be analyzed using 
Eq.~(\ref{eq39}) instead of Eq.~(\ref{velha}). 
We should mention that A. C. de la Torre, P. Catuogno and, S. Ferrando [7] had already
obtained the functions $\left< P_{1}P_{2} \right> - \left< P_{1} \right> \left< P_{2} \right>$ 
and $\left< Q_{1}Q_{2} \right> - \left< Q_{1} \right> \left< Q_{2} \right>$, which they call 
Quantum Covariance Functions (QCF), in a quite different context. They had proved that the QCF
vanishes if and only if the system is separable. This means that, as entanglement of a pair of 
identical particles means non-separability, the QCF does not vanish in Eq.~(\ref{eq39}) and we
do have an uncertainty relation that is different from Heisenberg's. 
  
In order to illustrate what we have discussed up to this point we now study
Eq.~(\ref{eq39}) for a particular 1-dimensional wave function:
\begin{equation}
	\Psi (x_{1},x_{2},t)=\int f(k_{1},k_{2}) \exp [i(k_{1}x_{1}+k_{2}x_{2}+\omega t)] 
	dk_{1}dk_{2}, \label{wave}
\end{equation}
where $\omega = \frac{E}{\hbar}$, $E$ is the energy of the system, $k_{1}$ and $k_{2}$ are the wave 
numbers of particles 1 and 2 respectively and 
\begin{displaymath}
f(k_{1},k_{2})=\frac{a}{(2\pi)^{3/2}} \left[ \exp \left( \frac{-a^{2}}{4}(k_{1}-k_{0})^{2} \right) 
\exp \left( \frac{-a^{2}}{4}(k_{2}+k_{0})^{2} \right) \right] + 
\end{displaymath}
\begin{equation}
\frac{a}{(2\pi)^{3/2}} \left[ \exp \left( \frac{-a^{2}}{4}(k_{1}+k_{0})^{2} \right) 
\exp \left( \frac{-a^{2}}{4}(k_{2}-k_{0})^{2} \right) \right].
\label{grande}
\end{equation}
Eq.~(\ref{grande}) represents a symmetric correlated pair of particles where 
$k_{1}+k_{2}=0$, that is, $k_{1}=-k_{2}=k_{0}$ or  $k_{1}=-k_{2}=-k_{0}$. 
Setting $t=0$ in Eq.~(\ref{wave}) we show that for  $k_{0}x_{1}\ll 1$ and $k_{0}x_{2}\ll 1$ that:
\begin{equation}
\left< x_{1} \right> = \left< x_{2} \right> = \left< p_{1} \right> = \left< p_{2} \right>=0,	 
\end{equation}
\begin{equation}
\left< x_{1}x_{2} \right> = \frac{k_{0}^{2}a^{4}}{2};  
\left< p_{1}p_{2} \right> = -2{\hbar}^{2}k_{0}^{2}.
\label{conta}	 
\end{equation}
Eq.~(\ref{conta}) shows that Eq.~(\ref{eq39}) should not be equal to the usual
Heisenberg uncertainty relation.
 
\section{CONCLUSION}

We have suggested here that when dealing with entangled systems of 
identical particles we should 
use {\em physical observables} (those that commute with all the permutation operators of the 
system) in whatever calculations we perform using the 
(anti-) symmetric wave function of the  system.

Applying the above assumption in the deduction of the uncertainty relation for a pair of 
entangled particles we have got (see Eq.~(\ref{eq39})) a relation which is more general 
than Heisenberg's uncertainty relation.
 
This new relation should not reduce to Heisenberg's relation when the particles are correlated. 

This generalized uncertainty relation suggests that, in conditional measurements, we can have states where 
$\Delta Q_{1}\Delta P_{1}<\frac{\hbar}{2}$. An example is the photon at the virtual slit in the experiment
of Kim and Shih. We can even get, at least theoretically, 
for example, a minimum dispersion for position without a divergence in the momentum dispersion.
 These possibilities we intend to explore further in the future.
 
\section*{ACKNOWLEDGMENTS}
The author acknowledges important suggestions, encouragement and useful discussions from
 C. O. Escobar and L. F. Santos. This research was supported by Funda\c{c}\~ao de Amparo \`{a} 
 Pesquisa do Estado de S\~ao Paulo. 
 
\section*{REFERENCES}
\begin{itemize}

\item[1.] Y. H. Kim and Y. Shih, ``Experimental realization of Popper's experiment: Violation of
the uncertainty principle?,'' {\em Found. Phys.} {\bf 29}, 1849 (1999).
\item[2.] K. R. Popper, {\em Quantum Theory and the Schism in Physics} 
(Hutchinson, London, 1982).
\item[3.] A. Einstein, B. Podolsky, and N. Rosen, ``Can quantum mechanical description of physical
reality be considered complete?,'' {\em Phys. Rev.} {\bf 47}, 777 (1935).
\item[4.] A. J. Short, ``Popper's experiment and conditional uncertainty relations,'' {\em Found.
Phys.} {\bf 14}, 275 (2001), quant-ph/0005063.
\item[5.] C. S. Unnikrishnan, ``Popper's experiment, uncertainty principle, signal locality and 
momentum conservation,'' {\em Found. Phys. Lett.} {\bf 13}, 197 (2000).
\item[6.] C. Cohen-Tannoudji, Bernard Diu, and Franck Lalo\"{e}, {\em Quantum Mechanics}, 
{\bf Vol. 2} (Hermann and John Wiley \& Sons, Paris, 1977).
\item[7.] A. C. de la Torre, P. Catuogno, and S. Ferrando, ``Uncertainty and nonseparability,''
{\em Found. Phys. Lett.} {\bf 2}, 235 (1989).
\end{itemize}

\end{document}